\documentclass[conference]{IEEEtran}

\usepackage{booktabs} % For formal tables
\usepackage{amsmath,amssymb,amsfonts,chemarrow,balance}
\usepackage{hyperref}
\usepackage{amsmath} 
\usepackage{graphicx}
\usepackage{subfigure}
\usepackage{multirow}
\usepackage{diagbox}
\usepackage{balance}
\usepackage{mathenv}
\usepackage{breqn}
\newcommand{\BfPara}[1]{{\noindent\bf#1.}\xspace\xspace}
\usepackage{url}
\usepackage{array}
\usepackage{stfloats}
\newcommand{\vs}[1]{{\vspace{-#1mm}}}

\usepackage{amsmath}
\usepackage{color}
\usepackage[linesnumbered,ruled,noend]{algorithm2e}
\usepackage{algpseudocode}
\usepackage{amsmath}
\usepackage{graphics}
\usepackage{amssymb}
\usepackage{bm}
\usepackage{rotating}
\usepackage{epsfig}

\SetKwRepeat{Do}{do}{while}%

\makeatletter

\newcommand{\bc}{{Bitcoin}\xspace}

\let\labelindent\relax
\usepackage[inline]{enumitem}

\def\equationautorefname~#1\null{(#1)\null}

\hypersetup{
	pdfpagemode=pagewidth,
	plainpages=false,
	colorlinks,
	urlcolor=blue,
	linkcolor=blue,
	citecolor=blue,
	bookmarksnumbered
}

\renewcommand{\baselinestretch}{0.94}

\begin{document}

\title{Exploring Spatial, Temporal, and Logical Attacks on the Bitcoin Network}

\author{\IEEEauthorblockN{Muhammad Saad$^\star$, Victor Cook$^\star$, Lan Nguyen$^\dagger$, My T. Thai$^\dagger$, and Aziz Mohaisen$^\star$}
\IEEEauthorblockA{ $^\star$University of Central Florida $^\dagger$ University of Florida   }
\IEEEauthorblockA{ \{saad.ucf, victor.cook\}@knights.ucf.edu, lan.nguyen@ufl.edu, mythai@cise.ufl.edu, mohaisen@cs.ucf.edu  }

}

% \IEEEoverridecommandlockouts
% \makeatletter\def\@IEEEpubidpullup{6.5\baselineskip}\makeatother
% \IEEEpubid{\parbox{\columnwidth}{
%     Network and Distributed Systems Security (NDSS) Symposium 2019\\
%     24-27 February 2019, San Diego, CA, USA\\
%     ISBN 1-891562-55-X\\
%     https://dx.doi.org/10.14722/ndss.2019.23xxx\\
%     www.ndss-symposium.org
% }
% \hspace{\columnsep}\makebox[\columnwidth]{}}
\maketitle

\begin{abstract}
In this paper, we explore the partitioning attacks on the Bitcoin network, which is shown to exhibit spatial bias, and  temporal and logical diversity. Through data-driven study we highlight: 
\begin{enumerate*}
    \item the centralization of Bitcoin nodes across autonomous systems, indicating the possibility of BGP attacks, 
    \item the non-uniform consensus among nodes, that can be exploited to partition the network, and
    \item the diversity in the Bitcoin software usage that can lead to privacy attacks.
\end{enumerate*}
Atop the prior work, which focused on spatial partitioning, our work extends the analysis of the Bitcoin network to understand the temporal and logical effects on the robustness of the Bitcoin network.

\end{abstract}

\section{Introduction}\label{sec:introduction}
The Bitcoin network consists of nodes that are connected in a peer-to-peer architecture. These nodes are geographically spread over the Internet, and they use a gossip protocol to exchange transactions and blocks. Ideally, Bitcoin nodes are expected to remain synchronized over the state of the blockchain in order to maintain a consistent view. Moreover, the decentralized and distributed network is considered safe against single point-of-failure. However, these assumptions, if challenged, may lead to system-wide vulnerabilities and partitioning attacks, which we explore in this paper.

In Bitcoin, partitioning attacks can be launched against a group of nodes that:
\begin{enumerate*}
    \item are geographically clustered withing an autonomous system (AS),
    \item have an outdated view of the blockchain, and
    \item are using a vulnerable software client infected with malware and bugs.  
\end{enumerate*}
As an outcome of each attack, an adversary can influence the key features and operations of Bitcoin including the publication of a block, the transaction confirmation, and the network size. Prior work~\cite{ApostolakiAL17}, on partitioning attacks focused on the spatial attack vector, indicating that Bitcoin nodes are vulnerable to BGP hijacks. In this paper, we validate their findings and present an up-to-date condition of the network. Additionally, and novel to our work, we present more optimized and cost effective attacks that extend into temporal and logical network frontiers.

\BfPara{Data Collection}
For this study, we crawled Bitnodes~\cite{BN18}, a service that maintains a persistent connection with all reachable nodes in the Bitcoin network. We used the information to develop another crawler, atop Bitnodes, to acquire data. The data contained information including the IP address, the latest block, and the software client of each node. We used the IP address of nodes to find their corresponding AS and organization. For our analysis, we use two datasets: a sample per minute and a sample per 10 minutes, respectively.%We ran two crawlers in parallel, each with a sampling time of 10 minutes and one minute respectively.  

\vs{2}
\section{Partitioning Attacks}\label{sec:PA}
% In the following, we describe the three forms of partitioning attacks that can be launched on Bitcoin. 

\begin{figure}[!t]
\begin{center}
\includegraphics[width=0.30\textwidth]{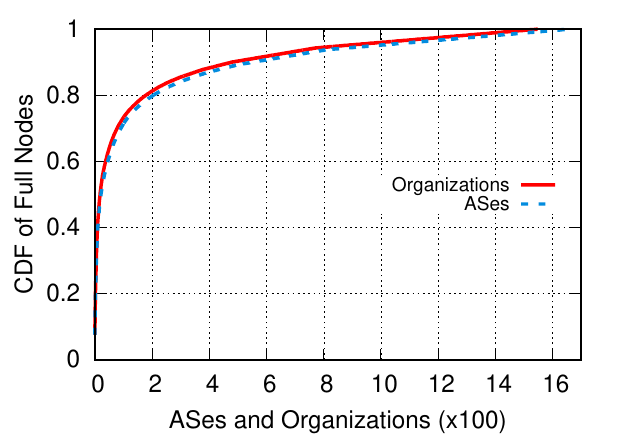}
\vs{3}
\caption{CDF of the Bitcoin full nodes in ASes and organizations.} 
\label{fig:asn}
\end{center}
\vs{2}
\end{figure}

\subsection{Spatial Partitioning} \label{sec:spp}
In spatial partitioning, an adversarial AS or organization can hijack BGP prefixes of a target AS that hosts a higher fraction of Bitcoin nodes and mining pools. As a result, it can hijack the Bitcoin traffic, isolate the mining power, or simply harm the reputation of the target AS. 

Prior work~\cite{ApostolakiAL17}, carried out in 2017, showed that 13 ASes hosted 30\% Bitcoin nodes while 50 ASes hosted 50\% Bitcoin nodes. In our analysis, started on February 28, 2018, we found that only 8 ASes host 30\% of Bitcoin nodes and 24 ASes host 50\% of Bitcoin nodes. At the organization level, we found that only 13 organizations host 50\% of the Bitcoin nodes. Among them, only two organizations host 65.7\% of Bitcoin hash rate, with the leading organization ({\em AliBaba}) having a 59.4\%  share of \bc the hash rate. This indicates that since 2017, Bitcoin has become more centralized. In~\autoref{fig:asn}, we plot the CDF of ASes and organizations that host Bitcoin full nodes, and in ~\autoref{tab:mp}, we present the top 5 mining pools along with their hash rate and distribution across ASes and organizations.  

\begin{figure*}[t]
\hfill
\begin{subfigure}[General trend of the network\label{fig:behindDays}]{\includegraphics[width=0.25\textwidth]{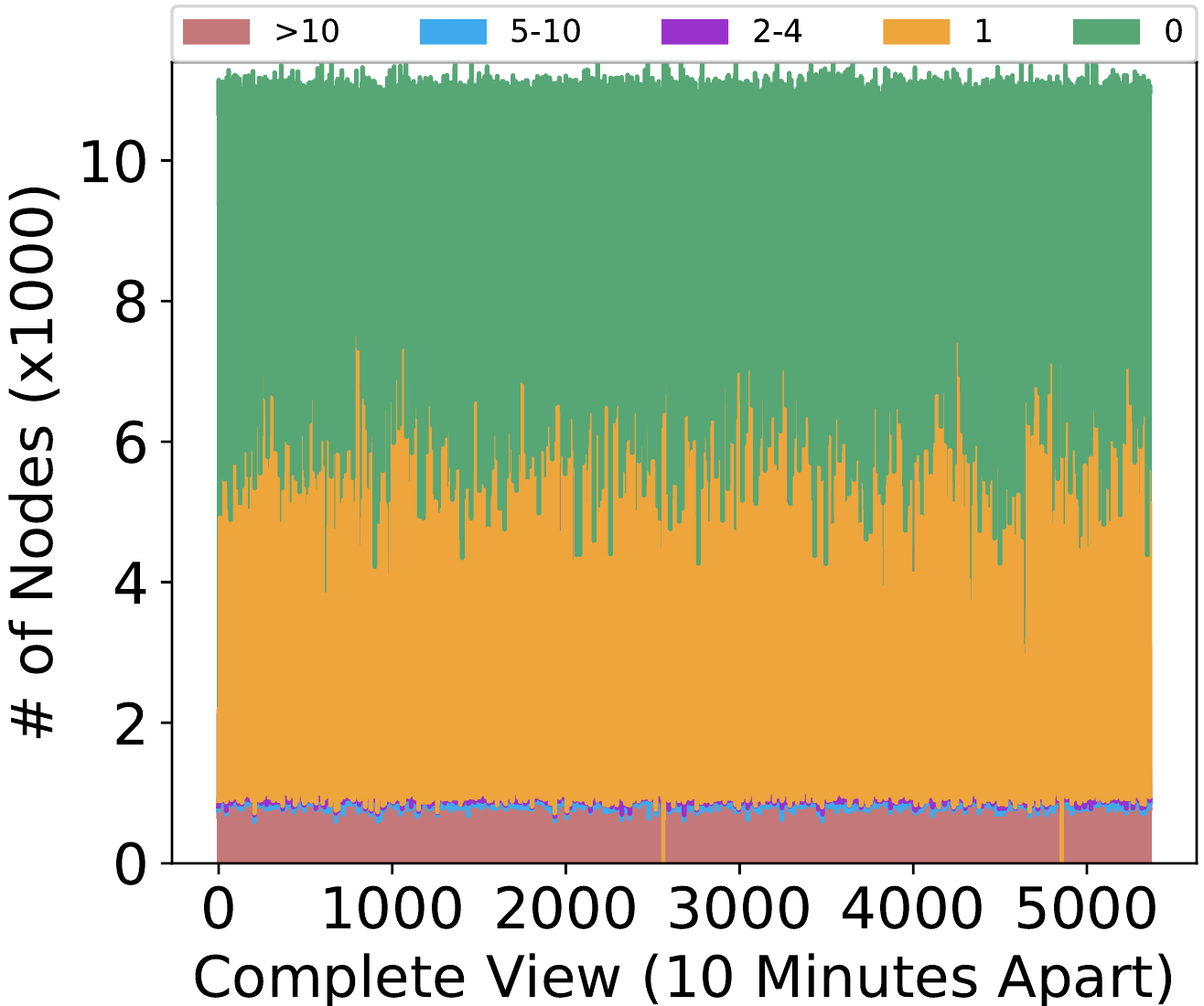}} 
\hfill
\end{subfigure}
\begin{subfigure}[One day snapshot \label{fig:behindone}]{\includegraphics[width=0.25\textwidth]{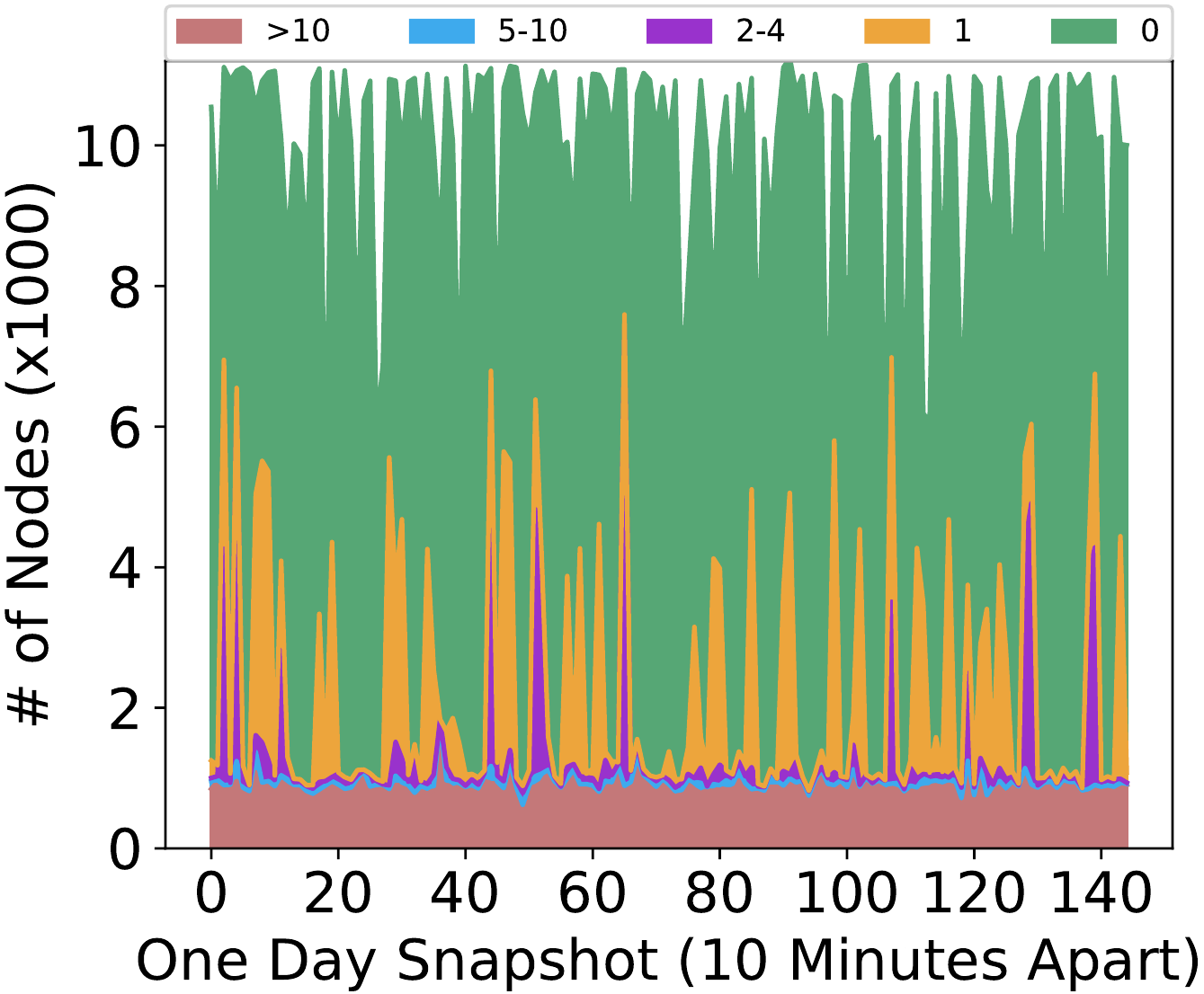}}
\hfill
\end{subfigure}
\begin{subfigure}[Consensus between block propagation \label{fig:per-minute}]{\includegraphics[width=0.25\textwidth ]{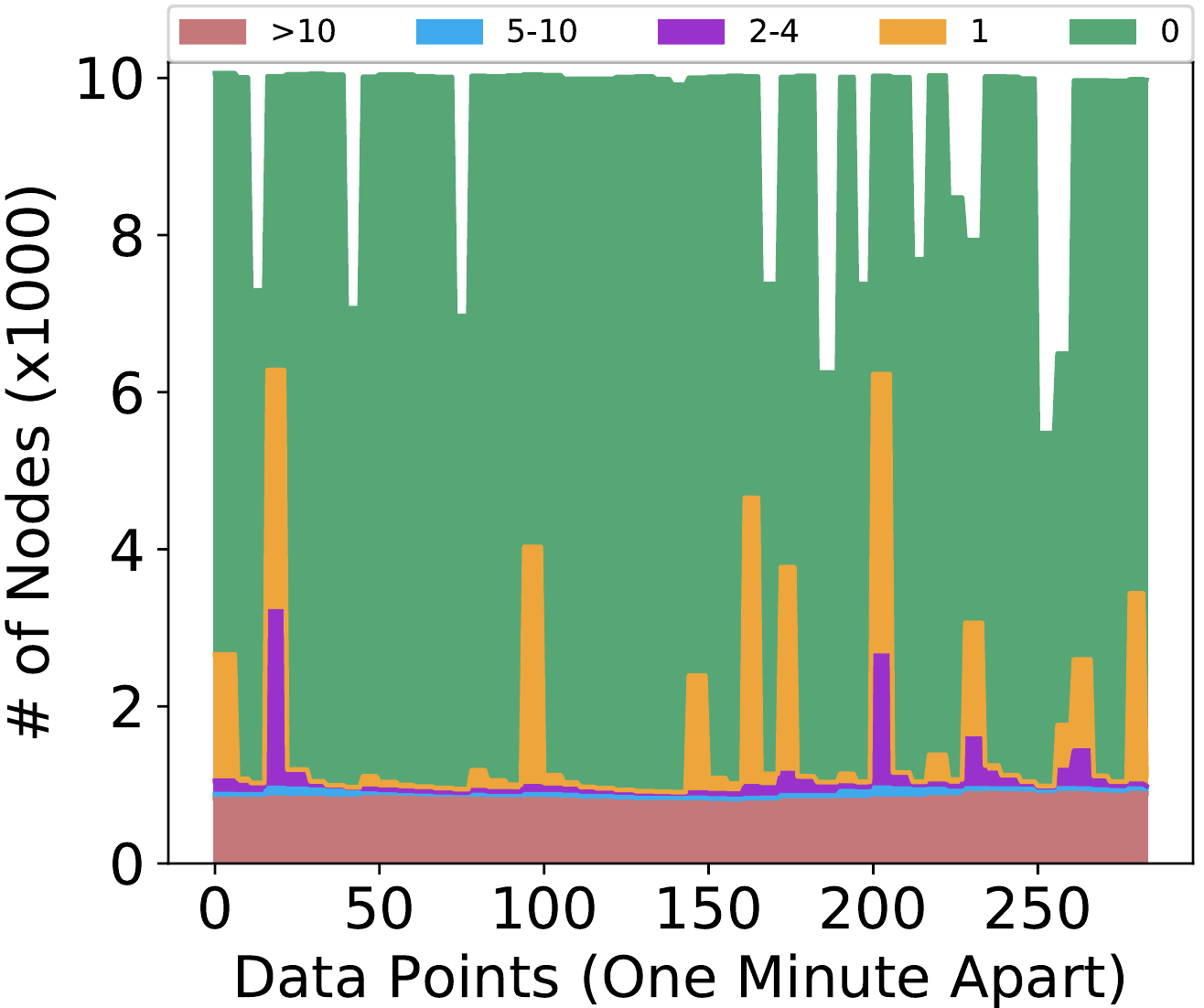}} 
\hfill
\end{subfigure}
\vs{4}
\caption{ Temporal consensus in Bitcoin network. Y-axis denotes number of nodes in 1000. In each figure, green region denotes the up-to-date blocks. Yellow region denotes 1 block behind. Purple, blue, and magenta regions represent nodes that are 2--4, 5--10, and $\geq$ 10 blocks behind respectively.~\autoref{fig:behindDays} shows the overall network, ~\autoref{fig:behindone}, shows a day (March 25) that offers greater attack opportunity, and ~\autoref{fig:per-minute} shows consensus pruning during 10 minutes. } 
\vs{3}
\label{fig:temporal}
\end{figure*}

\begin{table}[t]
\centering
\caption{Top 5 mining pools per hash rate, ASes and organization. Alibaba has a view of $\approx$ 60\% of the mining data. }
\label{tab:mp}
\scalebox{0.95}{
\begin{tabular}{l|c|c|c}
\hline
\multicolumn{1}{l|}{\textbf{Mining Pool}} & \textbf{H. Rate \% } & \textbf{ASes} & \textbf{Organizations} \\ \hline
{\multirow{2}{*}{BTC.com }}                                & {\multirow{2}{*}{25\%  }}                 & AS37963       & Hangzhou Alibaba       \\ 
                                           &                       & AS45102       & AliBaba (China)        \\ 
\multicolumn{1}{l|}{Antpool}              & 12.4\%                & AS45102       & AliBaba (China)        \\ 
\multicolumn{1}{l|}{ViaBTC}               & 11.7\%                & AS45102       & AliBaba (China)        \\ 
\multicolumn{1}{l|}{BTC.TOP}              & 10.3\%                & AS45102       & AliBaba (China)        \\ 
{\multirow{2}{*}{F2Pool}}                                    & {\multirow{2}{*}{6.3\%}}                 & AS45102       & AliBaba (China)        \\ 
                                           &                       & AS58563       & Chinanet Hubei         \\

%{\multirow{1}{*}{\textbf{Subtotal}}}                                    & {\multirow{1}{*}{65.7\%}}                 & ---      & ---       \\ \cline{1-4} 

{\multirow{1}{*}{12 others}}                                    & {\multirow{1}{*}{34.3\%}}                 & ---      & ---       \\ \hline

%{\multirow{1}{*}{\textbf{Total}}}                                    & {\multirow{1}{*}{100\%}}                 & ---      & ---       \\ \cline{1-4} 
\end{tabular}}
\vs{4}
\end{table}

Spatial partitioning can facilitate other major attacks including double-spending, consensus delay, eclipse attacks, and the 51\% attack. As shown in~\autoref{tab:mp}, if an attacker hijacks 3 ASes, he can isolate more than 60\% of the Bitcoin hash rate. This can be further extended by individually targeting ASes and hijacking BGP prefixes. To prevent spatial partitioning, node hosting should be spread across multiple ASes. This can resist the centralization and raise the attack cost.

\vs{2}
\subsection{Temporal Partitioning} \label{sec:tmp}
In temporal partitioning, a malicious miner can partition the network and force users into following a counterfeit blockchain. The objective of the attacker is the isolation and subversion of nodes that are behind the main chain by one or more blocks. The nodes can be behind the chain due to poor connectivity, low bandwidth, or network churn. 

An attacker with the information of vulnerable nodes can connect to them and feed them false blocks. To outline the network's vulnerability, we plot this temporal diversity of Bitcoin nodes in~\autoref{fig:temporal}, where  x-axis denotes a time-index for network observations (one observation every 10 minutes in \autoref{fig:behindDays} and \autoref{fig:behindone}, and one every minute in \autoref{fig:per-minute}). From~\autoref{fig:temporal}, we were able to make following observations.
\begin{enumerate*}
    \item Consensus pruning is not uniform across the network.
    \item Generally, a majority ($\approx 50$\%) remains synchronized. 
    \item 30--40\% nodes remain 1--4 blocks behind with respect to updated nodes.
      \item There are vulnerable moments in which up to 90\% of the network is 1--4 blocks behind. 
\end{enumerate*}

As shown in \autoref{fig:temporal}, nodes in the yellow and purple region are behind the network, and vulnerable to attacks. The attack recovery will require a fork in which all transactions will be reversed and UTXO sets will be updated. Standing out in our analysis is the observation that Bitcoin has a level of asymmetric vulnerability. With a market capitalization of $o(10^{11})$ USD and network configuration of $o(10^4)$ nodes, each full node is worth $o(10^7)$ USD.  However, the cost of disrupting the network is far less than the value being impacted.

Since temporal partitioning has not been studied before, therefore, effective countermeasures do not exist. However, we propose a simple yet effective scheme, called {\em BlockAware}~\cite{SaadCLTM}, which uses the expected block time to notify the node about its blockchain view with respect to the network. As part of our ongoing work, we are prototyping {\em BlockAware} over Bitcoin Core to defend against the temporal attacks.

\vs{2}
\subsection{Logical Partitioning} \label{sec:lgp}
To connect to the Bitcoin network, peers run a software called Bitcoin Core that implements protocols of the system. Bitcoin Core is an open source project that can be customized and updated to implement new rules and policies. Since 2009, there have been over 40 updates to Bitcoin Core, with the latest, v0.16.0 released in February 2018. 

~\autoref{tab:topsw} shows the distribution of Bitcoin software at the time of our data collection. We observed that 288 Bitcoin software variants are used by nodes. The latest version of Bitcoin Core, 0.16.0, is used by only 36\% of the nodes while 27\% use version 0.15.1.  The remaining 37\% of the network uses 286 different software clients.

\begin{table}[]
\centering
\caption{Top 5 software versions used by Bitcoin full nodes along with their release date, lag from the date of collection in days, and percentage of users. }
\label{tab:topsw}
\scalebox{0.99}{
\begin{tabular}{c|l|c|c|c}
\hline
\textbf{Index} & \textbf{Version} & \textbf{Release Date} & \textbf{Lag} & \textbf{Users \%} \\ \hline
1              & B. Core v0.16.0      & 02-26-2018  & 59          & 36.28\%           \\ 
2              & B. Core v0.15.1      & 11-11-2017   & 166          & 27.52\%           \\ 
3              & B. Core v0.15.0.1    & 09-19-2017   & 219          & 5.01\%            \\ 
4              & B. Core v0.14.2      & 06-17-2017   & 313          & 4.67\%            \\ 
5              & B. Core v0.15.0      & 04-22-2017   & 369          & 2.05\%            \\ \hline
\end{tabular}}
\vs{2}
\end{table}

Peer ``democracy'' in software selection has served well, but is vulnerable to attacks. An attacker can release a modified version of software client, contaminated with bugs and malware that can put the privacy of the user at risk. To obfuscate the true nature of the software client, and to gain confidence of the users, the attacker may also introduce useful features in his software that offer better performance. One example is Falcon, that provides faster connectivity and minimum delay during transaction propagation~\cite{zhang2007practical}. Falcon is not malicious, but it demonstrates the independence of peers to run a client that is not part of Bitcoin Core. Logical partitioning can compromise the privacy of a node running the malicious software version. It may expose the user to privacy risks and theft. 

Vulnerability to logical partitioning is due to the open network protocol. A central authority to regulate client participation would violate decentralization, and therefore, logical partitioning attacks remain a vulnerability to be considered.

\vs{2}
\section{Conclusion} \label{sec:conclusion}
In this paper, we explore the vulnerability of Bitcoin to spatial, temporal, and logical attacks. Our results show that over time, Bitcoin nodes have become more centralized among ASes, and therefore, more vulnerable to BGP attacks. Additionally, due to the diversity in consensus and the use of different software clients by different nodes, both temporal and logical partitioning attacks are possible. 

\BfPara{Acknowledgement:}This work is supported by Air Force Material Command award FA8750-16-0301.

% {\normalsize \bibliographystyle{acm}
% \bibliography{references,conf}}
\balance
\bibliographystyle{IEEEtranS}
\bibliography{references.bib,conf.bib}

\end{document}